\documentstyle[preprint,epsfig,aps]{revtex}
\begin{document}
\preprint{\vbox{\hbox{IFT-P.079/97}     \vspace{-.3cm}
                \hbox{IFUSP/P-1289}     \vspace{-.3cm}
                \hbox{FTUV/97-59}       \vspace{-.3cm}
                \hbox{IFIC/97-89}       \vspace{-.3cm}
                \hbox{hep-ph/9711499}}}

\title{Tests of Anomalous Quartic Couplings at the NLC}

\author{O.\ J.\ P.\ \'Eboli$^1$\thanks{Email: eboli@ift.unesp.br;
    $^\dagger$ Email: concha@axp.ift.unesp.br; $^\ddagger$ Email:
    mizuka@fma.if.usp.br} , M.\ C.\ Gonzalez-Garcia$^{1,2,\dagger}$,
    and J.\ K.\ Mizukoshi$^{3,\ddagger}$}

\address{$^1$ Instituto de F\'{\i}sica Te\'orica, 
              Universidade Estadual Paulista \\   
              Rua Pamplona, 145,
              01405--900 --  S\~ao Paulo, Brazil} 

\address{$^2$ Instituto de F\'{\i}sica Corpuscular --  IFIC/CSIC,
              Departament de F\'{\i}sica Te\`orica \\
              Universitat de Val\`encia, 46100 Burjassot, Val\`encia, Spain}

\address{$^3$ Instituto de F\'{\i}sica ,
              Universidade de S\~ao Paulo \\
              C.\ P.\ 66.318, 05315--970 --  S\~ao Paulo, SP, Brazil.}

\maketitle
\vskip -0.5cm
\begin{abstract} 
  
  We analyze the potential of the Next Linear $e^+e^-$ Collider to
  study anomalous quartic vector--boson interactions through the
  processes $e^+ e^- \rightarrow W^+W^-Z$ and $ZZZ$.  In the framework
  of $SU(2)_L \otimes U(1)_Y$ chiral Lagrangians, we examine all
  effective operators of order $p^4$ that lead to four-gauge-boson
  interactions but do not induce anomalous trilinear vertices. In our
  analysis, we take into account the decay of the vector bosons to
  fermions and evaluate the efficiency in their reconstruction. We
  obtain the bounds that can be placed on the anomalous quartic
  interactions and we study the strategies to distinguish the possible
  couplings.

\end{abstract}


\section{Introduction}

The impressive agreement of the Standard Model (SM) predictions for
the fermion--vector boson couplings with the experimental results has
been a striking confirmation of the $SU_L(2)\times U_Y(1)$ gauge
structure of the model in that sector \cite{ewwg}.  However, we still
lack the same accuracy tests for the structure of the bosonic sector.
If the gauge and symmetry breaking sectors are invariant under the
$SU_L(2) \times U_Y(1)$ gauge group, the structure of the triple and
quartic vector-boson is completely determined. Thus a detailed study
of these interactions can either confirm the local gauge invariance of
the theory or indicate the existence of new physics beyond the SM.

Presently, we have only started to probe directly the triple
gauge--boson couplings at the Tevatron \cite{teva,cgn} and LEP
\cite{lep} through the production of pairs of vector bosons. Notwithstanding,
the constraints on these couplings are still very loose. Future hadron
\cite{anohad} and $e^+e^-$, $e\gamma$, and $\gamma\gamma$
\cite{anoNLC} colliders will provide further information on these
couplings and improve significantly our knowledge of possible
anomalous gauge-boson interactions.

If the $SU(2)_L \otimes U(1)_Y$ symmetry of the model is to be
linearly realized, these studies of the triple gauge--boson couplings
will be able to furnish information on the gauge--boson four--point
functions provided that dimension 8 and higher anomalous operators are
suppressed. This is the case when the breaking of the $SU(2)_L \otimes
U(1)_Y$ symmetry takes place via the Higgs mechanism with a relatively
light elementary Higgs boson. If, on the other hand, no fundamental
light Higgs particle is present in the theory, one is led to consider
the most general effective Lagrangian which employs a nonlinear
representation of the broken $SU(2)_L \otimes U(1)_Y$ gauge symmetry.
In this case the relation between the structure of the three-- and
four--point functions of the gauge bosons does not hold already at
$p^4$ order, leaving open the question of the structure of the quartic
vector--boson interactions.

At present the only information on quartic gauge--boson interactions
is obtained {\it indirectly} as they modify the gauge--boson
two--point functions at one loop \cite{adri}.  The precise electroweak
measurements both at low energy and at the $Z$ pole, constrains the
quartic anomalous couplings to be smaller than $10^{-3}$--$10^{-1}$
depending on the coupling.
 
Direct studies of quartic vector--boson interactions cannot be
performed at the present colliders since the available centre-of-mass
energy is not high enough for multiple vector--boson production.  This
crucial test of the gauge structure of the SM will only be possible  at
the CERN Large Hadron Collider (LHC) through the reaction $pp
\rightarrow V_L V_L X$ \cite{deh,bdv,dhpru} or at the next linear
collider (NLC) through the processes $e^+ e^- \rightarrow VVV$
\cite{bhp,bela1,stir}, $e^+ e^- \rightarrow F F VV$ \cite{bhk}, $e^-
e^-\rightarrow F F V V $ \cite{cuypers}, $e\gamma \rightarrow V V F$
\cite{our:eg}, $\gamma\gamma \rightarrow VV$ \cite{bela2}, and $\gamma
\gamma \rightarrow VVV $ \cite{our:vvv}, where $V=$ $Z$, $W^\pm$ or
$\gamma$ and $F=$ $e$ or $\nu_e$.

In this work we analyze in detail the processes $e^+ e^- \rightarrow
W^+ W^- Z$ and $ZZZ$ in order to assess the potential of the NLC, with
and without polarized beams, to study anomalous quartic couplings of
vector bosons. These reactions will be the most important processes to
study the quartic gauge couplings at the NLC up to energies of the
order of 1 TeV, where the processes $e^+ e^- \rightarrow V V F F$
start to become important \cite{fbou}. We work in the framework of
chiral Lagrangians, and we study all $p^4$ operators that lead genuine
quartic gauge interactions, {\em i.e.} these operators do not give
rise to triple gauge--boson vertices, and consequently are not bounded
by the study of the production of gauge--boson pairs. We extend the
analysis of Ref.\ \cite{bela1} for the custodial $SU(2)_C$ conserving
operators taking into account realistic cuts and detection
efficiencies.  Moreover, we also study the non-conserving $SU(2)_C$
interactions.


\section{Theoretical Framework}

If the electroweak symmetry breaking is due to a heavy (strongly
interacting) Higgs boson, which can be effectively removed from the
physical low--energy spectrum, or to no fundamental Higgs scalar at
all, one is led to consider the most general effective Lagrangian
which employs a nonlinear representation of the broken $SU(2)_L
\otimes U(1)_Y$ gauge symmetry \cite{Appelquist}. The resulting chiral
Lagrangian is a non--renormalizable non--linear $\sigma$ model coupled
in a gauge--invariant way to the Yang--Mills theory.  This model
independent approach incorporates by construction the low-energy
theorems \cite{cgg}, that predict the general behavior of Goldstone
boson amplitudes irrespective of the details of the symmetry breaking
mechanism. Notwithstanding, unitarity implies that this low--energy
effective theory should be valid up to some energy scale smaller than
$4\pi v \simeq 3$ TeV \cite{lee}, where new physics would come into
play.

To specify the effective Lagrangian one must first fix the symmetry
breaking pattern. We consider that the system presents a global
$SU(2)_L \otimes SU(2)_R$ symmetry that is broken to $SU(2)_C$. With
this choice, the building block of the chiral Lagrangian, in the
notation of Ref.\ \cite{Appelquist}, is the dimensionless unimodular
matrix field $\Sigma(x)$, which transforms under $SU(2)_L \otimes
SU(2)_R$ as $(2,2)$:
\begin{equation}
\Sigma(x) ~=~ \exp\left(i \frac{\varphi^a(x) \tau^a}{v}\right) \; .
\end{equation}
The $\varphi^a$ fields are the would-be Goldstone fields and $\tau^a$
($a=1$, $2$, $3$) are the Pauli matrices.  The $SU(2)_L
\otimes U(1)_Y$ covariant derivative of $\Sigma$ is defined as
\begin{equation}
D_\mu \Sigma ~\equiv~ \partial_\mu \Sigma 
+ i g \frac{\tau^a}{2} W^a_\mu \Sigma -
i g^\prime \Sigma \frac{\tau^3}{2} B_\mu \; .
\end{equation}

The lowest-order terms in the derivative expansion of the effective
Lagrangian are
\begin{equation}
{\cal L}^{(2)} = \frac{v^2}{4} \hbox{Tr} \left [ \left ( D_\mu \Sigma \right )
^\dagger \left ( D^\mu \Sigma \right ) \right ]
+ \beta_1 g'^2\frac{v^2}{4} \left ( \hbox{Tr}
\left [ T V_\mu \right ] \right )^2
\; .
\label{lagran2}
\end{equation}
where we have introduced the auxiliary quantities $T \equiv\Sigma
\tau^3 \Sigma^\dagger$ and $V_\mu \equiv \left ( D_\mu \Sigma \right )
\Sigma^\dagger $ which are $SU(2)_L$--covariant and
$U(1)_Y$--invariant. Notice that $T$ is not invariant under $SU(2)_C$
custodial due to the presence of $\tau^3$.

The first term in Eq.\ (\ref{lagran2}) is responsible for giving mass
to the $W^\pm$ and $Z$ gauge bosons for $ v = ( \sqrt{2} G_F )^{-1} $.
The second term violates the custodial $SU(2)_C$ symmetry and
contributes to $\Delta\rho$ at tree level, being strongly constrained
by the low--energy data. This term can be understood as the low-energy
remnant of a high--energy custodial symmetry breaking physics, which
has been integrated out above a certain scale $\Lambda$.  Moreover, at
the one--loop level, this term is also required in order to cancel the
divergences in $\Delta\rho$, arising from diagrams containing a
hypercharge boson in the loop.  This subtraction renders $\Delta\rho$
finite, although dependent on the renormalization scale
\cite{Appelquist}.

At the next order in the derivative expansion, $D=4$, several
operators can be written down \cite{Appelquist}. We shall restrict
ourselves to those containing genuine quartic vector-boson
interactions, which are
\begin{eqnarray}
{\cal L}^{(4)}_4 &=& \alpha_4\left[{\rm{Tr}}
\left(V_{\mu}V_{\nu}\right)\right]^2
\label{eff:4}
\;, \\
{\cal L}^{(4)}_5 &=& \alpha_5\left[{\rm{Tr}}
\left(V_{\mu}V^{\mu}\right)\right]^2
\;, \\
{\cal L}^{(4)}_6 &=& \alpha_6 \; {\rm{Tr}}\left(V_{\mu}V_{\nu}\right)
{\rm{Tr}}
\left(TV^{\mu}\right){\rm{Tr}}\left(TV^{\nu}\right) \;, \\
{\cal L}^{(4)}_7 &=& \alpha_7\;{\rm{Tr}}\left(V_{\mu}V^{\mu}\right)
\left[{\rm{Tr}}\left(TV^{\nu}\right)\right]^2
\;, \\
{\cal L}^{(4)}_{10} &=& \frac{1}{2}
\alpha_{10}\left[{\rm{Tr}}\left(TV_{\mu}\right)
\;{\rm{Tr}}\left(TV_{\nu}\right)\right]^2
\; .
\label{eff:10}
\end{eqnarray}
In an arbitrary gauge, these Lagrangian densities lead to quartic
vertices involving gauge bosons and/or Goldstone bosons. In the
unitary gauge, these effective operators give rise to anomalous $ZZZZ$
(all operators), $W^+W^-ZZ$ (all operators except ${\cal
L}^{(4)}_{10}$), and $W^+ W^- W^+ W^-$ (${\cal L}^{(4)}_4$ and ${\cal
L}^{(4)}_5$) interactions.  Moreover, the interaction Lagrangians
${\cal L}^{(4)}_6$, ${\cal L}^{(4)}_7$, and ${\cal L}^{(4)}_{10}$
violate the $SU(2)_C$ custodial symmetry due to the presence of $T$ in
their definitions.  Notice that quartic couplings involving photons
remain untouched by the genuinely quartic anomalous interactions at
the order $D=4$.  The Feynman rules for the quartic couplings
generated by these operators can be found in the last article of Ref.\
\cite{Appelquist}.

In chiral perturbation theory, the $p^4$ contribution to the processes
$e^+ e^- \rightarrow W^+W^-Z$ and $ZZZ$ arises from the tree level
insertion of $p^4$ operators, as well as from one-loop corrections due
to the $p^2$ interactions, which renormalize the $p^4$ operators
\cite{Appelquist}. However, the loop corrections to the scattering
amplitudes are negligible in comparison to the $p^4$ contributions for the 
range of values of the couplings and center--of--mass energies
considered in this paper. Therefore, numerically, our analysis is
consistent even though we neglected the loop corrections and kept only
the tree--level $p^4$ contributions.


\section{Limits on Quartic Couplings}

In order to study the quartic couplings of vector bosons we analyzed
the processes
\begin{eqnarray}
        e^+ e^- &&\rightarrow W^+ W^- Z
\label{WWZ}
\;\; , \\
        e^+ e^- &&\rightarrow Z Z Z
\label{ZZZ}
\;\; ,
\end{eqnarray}
which may receive contributions from anomalous $WWZZ$ and $ZZZZ$
interactions.  We included in our calculations all SM and anomalous
contributions that lead to these final states. Therefore, we
consistently considered the effect of all interferences between the
anomalous and SM amplitudes. The scattering amplitudes were generated
using Madgraph \cite{madg} in the framework of Helas
\cite{hellas}, with the anomalous couplings arising from the Lagrangians
(\ref{eff:4}-\ref{eff:10}) being implemented as Fortran routines. 
Moreover, we include in our calculation the $W$'s and
$Z$'s decays taking into account the gauge boson widths, spin
structures, and correlations of the scattering amplitude.

\begin{figure}
\begin{center}
\mbox{\epsfig{file=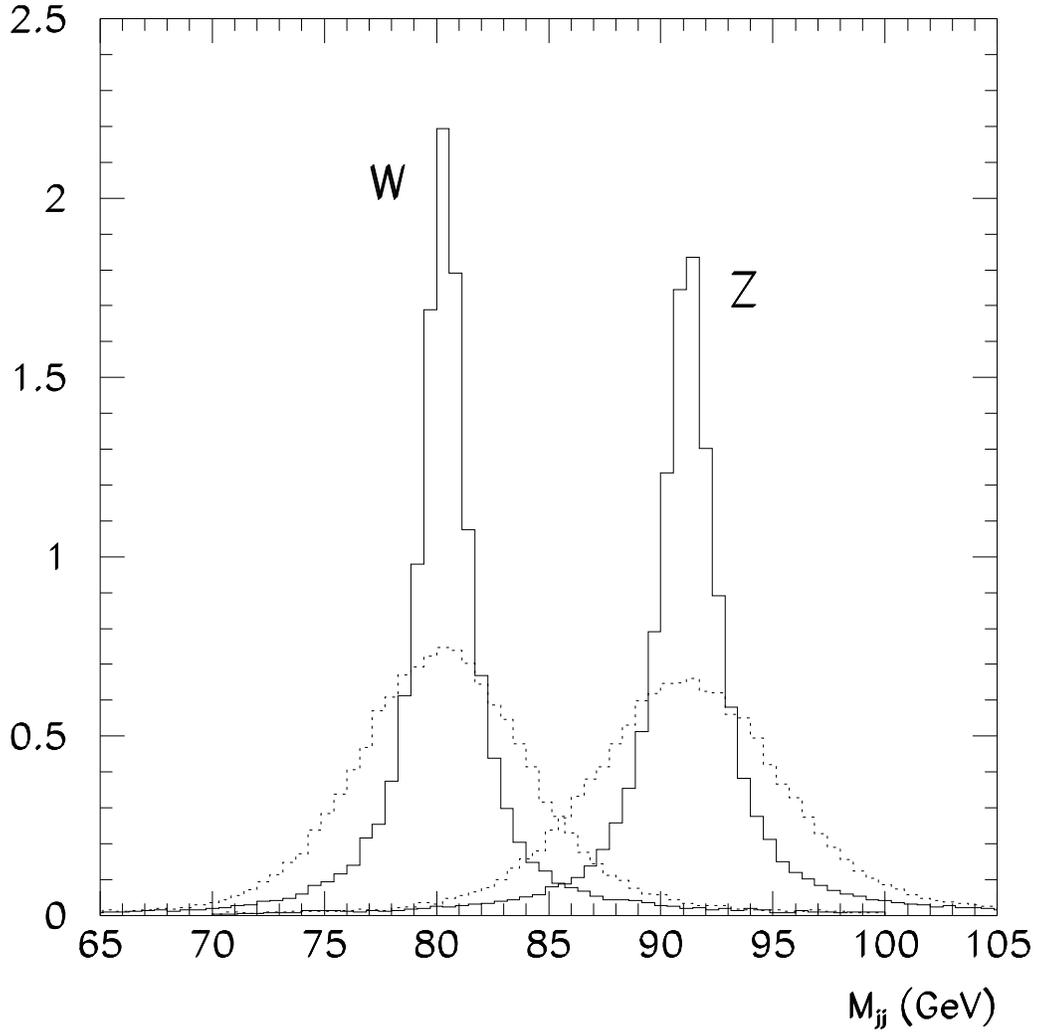,width=1\textwidth}}
\end{center}
\caption{Reconstructed invariant mass distribution for a jet pair
from $W$ and $Z$ decays. The full line only includes the effect of
the finite width while the dashed line contains also the effect of the
smearing due to the experimental resolution.}
\label{f:smear}
\end{figure}

We required the visible final state fermions to be in the rapidity
region $|\eta | < 3$ and separated by $ \Delta R =\sqrt{\Delta \eta^2
+ \Delta \phi^2} > 0.7$.  Furthermore, we also folded in the
experimental resolution factors associated to the electromagnetic and
hadronic calorimetry:
\begin{eqnarray}
\left. \frac{\delta E}{E}\right|_{em} &=& \frac{0.12}{\sqrt{E}} \oplus 0.01
\hbox{  eletromagnetic} \;\; ,
\\
\left.\frac{\delta E}{E}\right|_{had} &=& \frac{0.25}{\sqrt{E}} \oplus 0.02
\hbox{  hadronic} \;\; .
\end{eqnarray}
The momentum carried out by neutrinos was obtained using
energy--momentum conservation after smearing the momenta of final
state quarks and charged leptons. As illustration, we show in Fig.\
\ref{f:smear} the effect of the smearing on the reconstructed
difermion invariant masses for $W$'s and $Z$'s.

Difermion final states ($jj$, $\ell^+\ell^-$, $\ell^\pm \nu$, and
$\nu\nu$; with $\ell=e,\mu$) were identified as being a $W$ or a $Z$
provided their invariant masses after the smearing were in the range
\cite{vernon}, respectively,
\begin{equation}
\left [ 0.85 M_W~,~ \frac{1}{2} (M_W+M_Z)\right ]
\;\;\;\;,\;\;\;\;
\left [ \frac{1}{2} (M_W+M_Z) ~,~ 1.15 M_Z \right ] \;\; .
\label{cond:wz}
\end{equation}

In what follows we present our results for two different
center--of--mass energies, 500 GeV and 1 TeV, assuming an integrated
luminosity of 100 fb$^{-1}$ for both energies. We also study the
impact of a 80\% left--handed and 80\% right--handed polarized
electron beam while keeping the positron beam unpolarized.


\subsection{$e^+ e^- \rightarrow W^+ W^- Z$}

We identified $W^+W^-Z$ events through the topologies $6j$,
$4j+2\ell$, $4j+2\nu$, and $4j+\ell+\nu_\ell$, requiring two difermion
systems with invariant masses compatible with the $W$ mass --- see
Eq.\ (\ref{cond:wz})--- and one difermion system with an invariant
mass consistent with it being a $Z$. We show in Table \ref{t:wwz} the
fraction of $WWZ$ events that are reconstructed as $WWZ$ and $ZZZ$ for
center--of--mass energies of $0.5$ and $1$ TeV. It is interesting to
notice that the reconstruction probabilities are basically independent
of the $e^-$ polarization. Furthermore, the fraction of $WWZ$ events
reconstructed as $ZZZ$ generates a background for the study of
anomalous couplings in $ZZZ$ production.

The cross section for $W^+W^-Z$ ($ZZZ$) is a quadratic function of the
anomalous couplings $\alpha_i$, {\em i.e.}
\begin{equation}
\sigma_{\text{tot}} = \sigma_{\text{sm}} + \sum_i \alpha_i \; 
\sigma^{\alpha_i}_{\text{int}}
+ \sum_{ij} \alpha_i \alpha_j \; \sigma^{\alpha_i \alpha_j}_{\text{ano}} \; ,
\label{base}
\end{equation}
where $\sigma_{\text{sm}}$ stands for the SM cross section and
$\sigma^{\alpha_i}_{\text{int}}$ ($\sigma^{\alpha_i
\alpha_j}_{\text{ano}}$) is the interference (pure anomalous)
contribution.  In Table \ref{sig:wwz}, we present our results after
the cuts on $\eta$ and $\Delta R$, but before  $W$ and $Z$
identification. Therefore, these results should be multiplied by the
efficiencies given in Table \ref{t:wwz}. Notice that there are only
two independent Lorentz invariant structures for the $WWZZ$ vertices
at $p^4$ order, which implies that the couplings $\alpha_5$ and
$\alpha_7$ ($\alpha_4$ and $\alpha_6$) give rise to identical
contributions to $\sigma^{\alpha_i}_{\text{int}}$ and
$\sigma^{\alpha_i \alpha_j}_{\text{ano}}$ in $W^+W^-Z$
production. From this table we can witness that the SM contributions are
a slowly varying functions of the center--of--mass energy, while the
anomalous contributions grow rapidly, as one could naively
expect. Moreover, the SM background can be efficiently reduced using
right--handed electrons as this polarization eliminates almost
completely the contribution where the $W^-$ couples directly to the
$e^-$ fermion line.

In order to quantify the effect of the new couplings, we defined the
statistical significance $S$ of the anomalous signal
\begin{equation} 
S = \frac{|\sigma_{\text{tot}} -
\sigma_{\text{sm}}|}{\sqrt{\sigma_{\text{sm}}}} \; 
\sqrt{{\cal L}\epsilon} \; ,
\label{sig}
\end{equation} 
which can be easily evaluated using the parametrization (\ref{base})
with the coefficients given in Table \ref{sig:wwz}. In the above
expression $\epsilon$ stands for the detection efficiencies presented
in Table \ref{t:wwz}.

Table \ref{l:wwz} contains the values of the quartic anomalous
couplings that lead to an increase in the total number of events
smaller than $3\sigma$, assuming an integrated luminosity of 100
fb$^{-1}$ and that only one anomalous coupling is non-vanishing. These
limits were obtained combining events reconstructed as $W^+W^-Z$ from
all the topologies. It is interesting to notice that having
right--handed polarized electrons improves the bounds in 20--30\% with
respected to the results for unpolarized beams, while the use of
left--handed electrons weakens the limits.  This result is in
agreement with Ref.\ \cite{bela1}.  Moreover, the bounds improve as
the center--of--mass energy increases since the anomalous
contributions grow with energy.  In general, more than one anomalous
coupling might be non-vanishing.  In this case the correlation among
the anomalous couplings can be easily taken into account using the
full expression of Eq.\ (\ref{base}) and Table \ref{sig:wwz}.

In order to discriminate between the different couplings we studied
the kinematical distributions of the final gauge bosons.  Figure
\ref{f:wwz} displays the $W^+W^-$ invariant mass spectrum and the
$p_T$ distribution of the $Z$ in the $W^+W^-Z$ production with
unpolarized beams at $\sqrt{s}=500$ GeV. We plotted in this figure the
standard model prediction (dotted line) as well as the predictions for
$\alpha_4 = 0.61$ (dashed line) and $\alpha_5 = 0.38$ (solid line),
which are the values that lead to a $3\sigma$ signal in the total
number of events for unpolarized beams. As we can see, the $W^+W^-$
invariant mass distribution for $\alpha_4$ presents  a larger
contribution at small values of the $W^+W^-$ invariant mass, while
$\alpha_5$ gives rise to more events with larger invariant masses. In
principle we can use this distribution not only to distinguish the
anomalous couplings, but also to increase the sensitivity to the
signal. However, this can only be accomplished with higher integrated
luminosity. On the other hand the $p_T$ distributions of the $Z$ are
very similar in the SM and in presence of the anomalous couplings, being the
only difference the larger number of events in the
latter case.


\subsection{$e^+ e^- \rightarrow Z Z Z$}

The production cross section for $ZZZ$ final states is much smaller
than the one for $W^+W^-Z$, and consequently just a few fermionic
topologies can be used to identify these events. We considered only
the final states $6j$, $4j+2\ell$, and $4j+2\nu$, and required three
difermion systems with invariant masses compatible with the $Z$ one
according to the prescription given in Eq.\ (\ref{cond:wz}). We
present in Table \ref{t:zzz} the efficiency for the reconstruction of
the $ZZZ$ final state for the above topologies and center--of--mass
energies of 0.5 and 1.0 TeV. Analogously to the $W^+W^-Z$ case, these
efficiencies are independent of the polarization of the $e^-$.

Table \ref{sig:zzz} contains the values of $\sigma_{\text{sm}}$,
$\sigma^{\alpha_i}_{\text{int}}$, and
$\sigma^{\alpha_i\alpha_j}_{\text{ano}}$ for $ZZZ$ production, taking
into account the $\eta$ and $\Delta R$ cuts, but not the
reconstruction efficiencies. At $p^4$ order in chiral perturbation
theory, all the anomalous interactions are proportional to each other
since there is only one possible Lorentz structure for the vertex
which is multiplied by $\alpha_4+\alpha_5 + 2( \alpha_6 + \alpha_7 +
\alpha_{10})$.  Therefore, we only present the results for $\alpha_4$,
being straightforward the generalization to the other cases.  From
this table we can see that most of the reconstructed $ZZZ$ events will
be observed in the $6j$ and $4j+2\nu$ topologies. Furthermore, the
largest anomalous contribution comes from
$\sigma^{\alpha_i\alpha_j}_{\text{ano}}$, being the interference with
the SM of the same order of the SM contribution but with the opposite
sign.  Analogously, in $W^+W^-Z$ production, the anomalous
contributions grow substantially with the increase of the
center--of--mass energy, while the SM cross section decreases
slightly.

We present in Table \ref{l:zzz} the $3\sigma$ allowed for genuinely
quartic couplings that can be obtained from the non observation of
deviations from the SM in $ZZZ$ production. In our analysis of the
$6j$ topology, we included as background the $6j$ events coming from
$W^+W^-Z$ that are identified as $ZZZ$.  Despite the reduced number of
events in the $ZZZ$ channel, the bounds on the quartic couplings are
at least a factor of 2 better than the ones drawn from the $W^+W^-Z$
channel due to the smaller size of the background. On the other hand,
contrary to the $W^+W^-Z$ channel beam polarization does not lead to a
substantial improvement on the attainable limits.

\begin{figure}
\begin{center}
\mbox{\epsfig{file=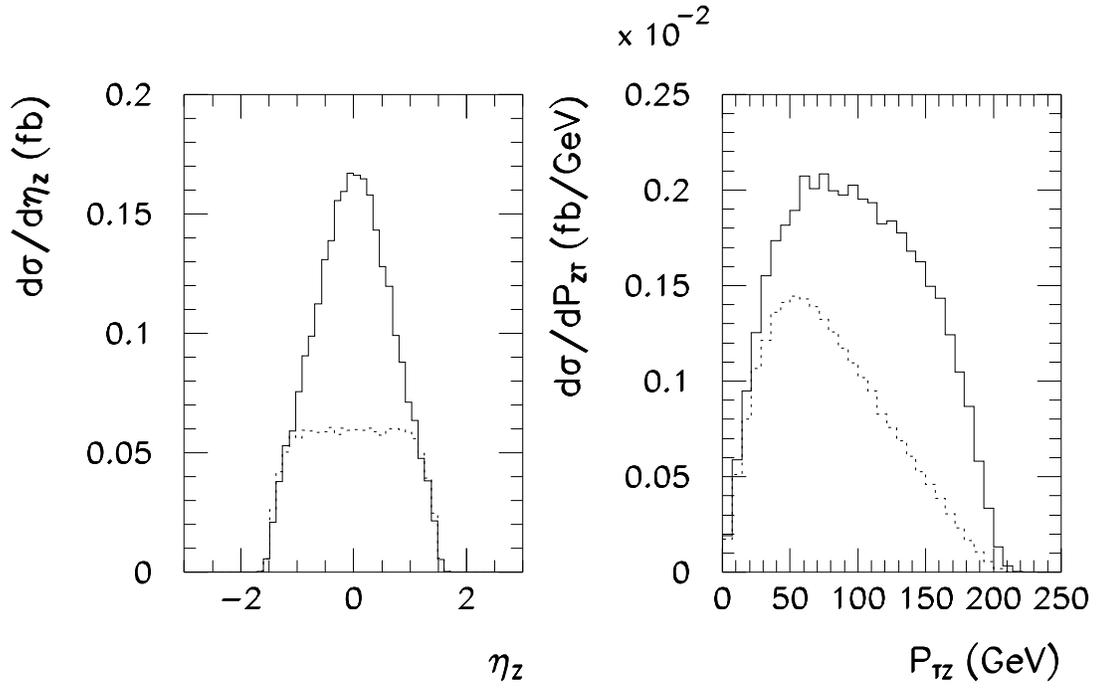,width=1\textwidth}}
\end{center}
\caption{Pseudorapidity and transverse momentum $Z$ distributions 
  in unpolarized $ZZZ$ production at $\protect\sqrt{s} = 500$ GeV.
  The dotted line stands for the SM result, while the solid 
  line represent the case $\alpha_4 =0.24$.  }
\label{f:zzz}
\end{figure}

We display in Fig.\ \ref{f:zzz} the pseudorapidity and transverse
momentum distribution of the $Z$'s in unpolarized $ZZZ$ production at
$\sqrt{s} = 500$ GeV. As we can see, the anomalous quartic
interactions leads to more centrally produced $Z$'s (smaller
$|\eta_z|$) which have a slightly harder $p_T$ spectrum. However, the
number of reconstructed events is not large enough to allow the use of
cuts to enhance the anomalous contributions.


\section{Discussion and Conclusions}
 
$W^+W^-Z$ and $ZZZ$ are the best channels for direct study of quartic
gauge--boson couplings in $e^+e^-$ colliders with center--of--mass
energies up to 1 TeV. At higher energies the most $e^+ e^- \rightarrow
W^+ W^- f \bar{f}$ becomes important process
\cite{bhk}. We showed in this work that the NLC will be able to
uncover the existence of anomalous quartic couplings of the order
${\cal O}(10^{-1})$ for center--of--mass energies up to 1 TeV and an
integrated luminosity of 100 fb$^{-1}$; see Tables \ref{l:wwz} and
\ref{l:zzz}. Despite these limits being weaker or of the order of the
present indirect bounds \cite{adri}, the above processes will provide
a {\em direct} test of the quartic interactions among the electroweak
gauge bosons. We have also shown that the use of a right--handed
polarized electron beam leads to better limits on the anomalous
interactions from the $W^+W^-Z$ production due to the substantial
reduction of the SM backgrounds.

\begin{figure}
\begin{center}
\mbox{\epsfig{file=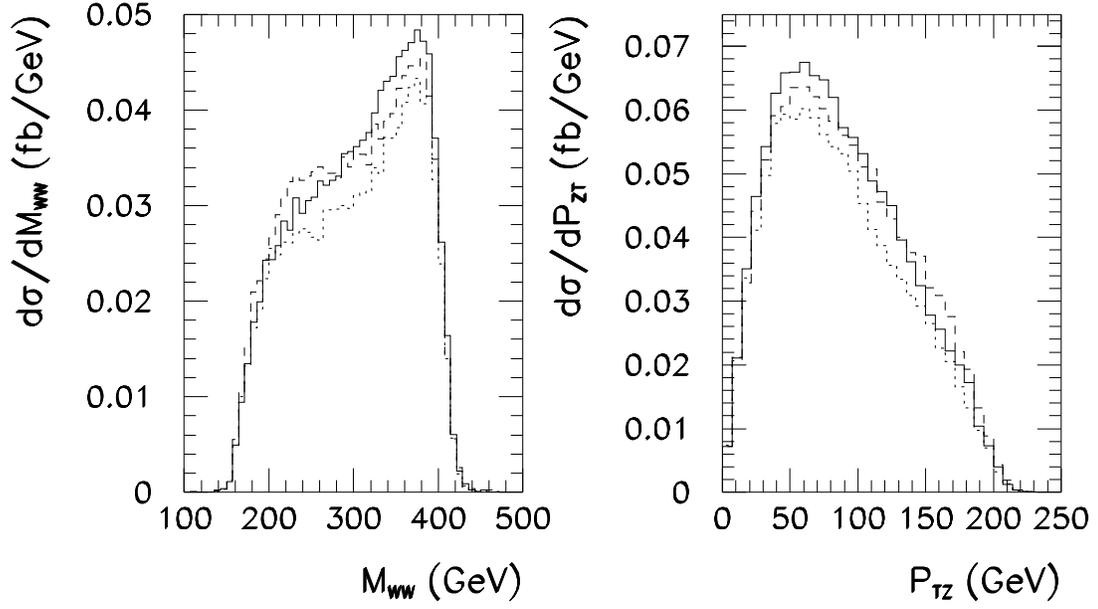,width=1\textwidth}}
\end{center}
\caption{ $W^+W^-$ invariant mass and $p_{TZ}$ distributions for
  unpolarized $W^+W^-Z$ production at $\protect\sqrt{s} = 500$ GeV.
  The dotted line stands for the SM result, while the solid (dashed)
  line represent the case $\alpha_5 = 0.38$ ($\alpha_4 = 0.61$).  }
\label{f:wwz}
\end{figure}

It is also important to devise a strategy to disentangle the anomalous
couplings in case a departure from the SM prediction is observed. In
$W^+W^-Z$ production, the analysis of the $W^+W^-$ invariant mass
distribution, see Fig.\ \ref{f:wwz}, can be used to distinguish
between the two possible structures for the $WWZZ$ vertex, one
associated to $\alpha_{4,6}$ and the other related to $\alpha_{5,7}$.
However, we are still left some two possibilities in both cases. At
this point it is important to use the information from the $ZZZ$
reaction, because the $SU(2)_C$ violating interactions leads to a much
larger excess of events for the same value of the anomalous coupling,
due to the coupling structure $\alpha_4+\alpha_5 + 2( \alpha_6 +
\alpha_7 + \alpha_{10})$. Therefore, the combination of the $W^+W^-$
distribution and the total number of events in both reactions are a
powerful tool to separate the effects of the different anomalous
couplings provided there is enough statistics. Moreover, the
comparison between the $W^+W^-Z$ event rates for different
polarizations can also be used to further distinguish between the
couplings $\alpha_{4,6}$ and $\alpha_{5,7}$, since the latter are less
sensitive to the electron polarization. Finally the anomalous coupling
$\alpha_{10}$ has the distinguished characteristics of modifying only
the $ZZZ$ production.


\acknowledgments

We would like to thank S.\ F.\ Novaes for discussions. M.\ C.\
Gonzalez--Garcia is grateful to the Instituto de F\'{\i}sica Te\'orica
from Universidade Estadual Paulista for its kind hospitality.  This
work was supported by FAPESP (Brazil), CNPq (Brazil), DGICYT (Spain)
under grant PB95--1077, and by CICYT (Spain) under grant AEN96--1718.

\medskip

\noindent{\bf Note added:} During the write up of this work we became aware
of a similar study performed by T.\ Han, H.-J.\ He, and C.-P.\ Yuan
\cite{han}.





\begin{table}
\begin{displaymath}
\protect
\begin{array}{||c|c|c||}
\hline\hline
\mbox{Topology}    &  F_{WWZ} \%  & F_{ZZZ} \%
\\
\hline \hline
6j                 &  64./52.         &  0.8/1.2
\\
4j+2\ell           &  66./55.         &  0.8/1.2
\\
4j+2\nu            &  28./8.          &  0.2/0.2
\\
4j+\ell\;\nu_\ell   &  20./5.          &  0./0.
\\
\hline\hline
\end{array}
\end{displaymath}
\caption{Fraction of $WWZ$ events that are reconstructed as $WWZ$ and
$ZZZ$ for several topologies and center--of--mass energies of $0.5$/$1$
TeV. }
\label{t:wwz}
\end{table}


\begin{table}[hbt]
\begin{center}
\begin{tabular}{||c|c|c|c|c|c|c|c||}
Topology &  &{ $\sigma_{\text{sm}}$ (fb)} &  
{ $-\sigma_{\text{int}}^{\alpha_4}$ (fb)}&  
{ $\sigma_{\text{ano}}^{\alpha_4\alpha_4}$ (fb)}& 
{ $\sigma_{\text{int}}^{\alpha_5}$ (fb)}& 
{ $\sigma_{\text{ano}}^{\alpha_5\alpha_5}$ (fb)} & 
{ $\sigma_{\text{ano}}^{\alpha_4\alpha_5}$ (fb)} \\ 
\hline
\hline
  & $\sqrt{s}/$TeV & 0.5 (1)  & 0.5 (1) & 0.5 (1)  &
  0.5 (1)  &  0.5 (1) &  0.5 (1)\\
 \hline
 & unpol & 7.41 (8.09)  & 0.12 (0.12) & 2.42 (6.30)  & 0.47 (0.21)  
& 4.53 (14.89)  &  1.18 (-4.71)\\ 
$6j$ & pol $-$ & 13.4 (14.59)  & 0.0 (0.0)  &2.81 (7.38)  & 0.57 (0.) 
          &5.26 (17.36) & 1.37 (-5.52) \\
& pol $+$ & 1.61 (1.60)  & 0.22 (0.24) & 2.0 (5.2) & 0.34 (0.38)
         & 3.77 (12.4)  & 0.99 (-3.93)\\
\hline
& unpol & 0.74 (0.80)  & 0.01 (0.01)    & 0.24 (0.63) 
& 0.01 (0.02) & 0.45 (1.48)  & 0.12(-0.47)\\
$4 j + 2 \ell$ &  pol $-$  &1.32 (1.43)    & 0.0 (0.0)  & 0.28 (0.72)& 
0.05 (0.0) & 0.52 (1.73)   & 0.14 (-0.53) \\
& pol $+$ & 0.159 (0.163) & 0.023 (0.023) & 0.20 (0.52) & 0.03 (0.04)
         & 0.375 (1.23) & 0.01 (-0.39)\\
\hline 
& unpol & 3.06 (2.88)  & 0.01 (0.09) & 0.94 (3.68) 
& 0.22 (0.16) & 2.09 (6.87) & 0.66 (-2.73) \\ 
 $4j+2\nu$    & pol $-$ & 5.44 (5.14)     & 0.0 (0.05) &1.09 (4.30)  
                  &0.29 (0.12) &2.44 (7.95) & 0.76 (-3.21) \\
& pol $+$ & 0.65 (0.58) &  0.06 (0.15)& 0.78 (3.08) 
                  & 0.15 (0.18) & 1.74 (5.7)  & 0.55(-2.27)\\         
\hline
 & unpol & 5.95 (6.38) & 0.12 (0.11)  & 1.97 (7.31) & 0.39
(0.48) & 3.45 (19.74)  & 0.98 (-3.63) \\
 $4 j + \ell\; \nu_\ell$   & pol $-$ & 10.63 (11.6) & 0.0 (0.0) &2.29 (8.60)  
                   & 0.49 (0.43) & 4.03 (23.12)  & 1.14(-4.27) \\
   &  pol $+$ & 1.28 (1.28)   & 0.2 (0.25) & 1.64 (6.14) 
                   & 0.26 (0.56) & 2.89 (16.4) & 0.81 (-3.04) \\
\end{tabular}
\end{center}
\caption{Values for the standard model, pure anomalous and
  interference cross sections (see Eq.\ \protect\ref{base})
for the $W^+W^-Z$ production and several center--of--mass energies 
and $e^-$ polarizations. The pol $-$ (pol $+$) lines correspond to
80\% left--handed (right--handed) electron beam polarization.}
\label{sig:wwz}
\end{table}


\begin{table}
\begin{displaymath}
\protect
\begin{array}{||c|c|c|c||}
\hline\hline
\sqrt{s} \mbox{ GeV}& e^- \mbox{polarization (\%)}&\alpha_{4,6}  &\alpha_{5,7}
\\
\hline \hline
 500   &           0    &   (-0.56,0.61)   & (-0.48, 0.38)
\\
 500   &         - 80    &   (-0.63,0.63)   & (-0.51,0.40)  
\\
 500   &           80    &   (-0.39,0.49)   & (-0.37,0.28)  
\\
1000   &           0    &   (-0.38,0.40)   & (-0.26,0.25)
\\
1000   &         - 80    &   (-0.43,0.43)   & (-0.28,0.28)      
\\
1000   &          80    &   (-0.27,0.32)   & (-0.20,0.17)      
\\
\hline\hline
\end{array}
\end{displaymath}

\caption{3$\sigma$ allowed values of the quartic anomalous couplings 
obtained from the reaction $e^+e^- \rightarrow W^+W^-Z$. 
 }
\label{l:wwz}
\end{table}


\begin{table}
\begin{displaymath}
\protect
\begin{array}{||c|c||}
\hline\hline
\mbox{Topology}    &  F_{ZZZ} \%
\\
\hline \hline
6j                 &  59. / 54.
\\
4j+2\ell           &  62. / 58.
\\
4j+2\nu            &  25. / 7.
\\
\hline
\hline
\end{array}
\end{displaymath}
\caption{Fraction of $ZZZ$ events that are reconstructed as
$ZZZ$ for several topologies and center--of--mass energies of $0.5$/$1$
TeV. }
\label{t:zzz}
\end{table}


\begin{table}[hbt]
\begin{center}
\begin{tabular}{||c|c|c|c|c||}
Topology &  &{ $\sigma_{\text{sm}}$ (fb)} &  
{ $-\sigma_{\text{int}}^{\alpha_4}$ (fb)}&  
{ $\sigma_{\text{ano}}^{\alpha_4\alpha_4}$ (fb)}
 \\ 
\hline
\hline
  & $\sqrt{s}/$TeV & 0.5 (1)  & 0.5 (1) & 0.5 (1)  
\\
 \hline
  & unpol  & 0.163 (0.145) & 0.169 (0.119) &2.83 (8.35)  
\\ 
$6 j$  & pol $-$  & 0.236 (0.21) & 0.224 (0.158) & 3.32 (9.77)  
\\
  &  pol $+$ &  0.090 (0.081) & 0.11 (0.082) & 2.38 (6.98)
\\
\hline
 & unpol  & 0.049 (0.043) & 0.044 (0.034)  & 0.846 (2.49) 
\\
  $4 j+ 2 \ell$ & pol $-$   & 0.07 (0.063)  & 0.057 (0.047)  & 0.985 (2.90) 
\\
  &  pol $+$ & 0.027 (0.024) & 0.03 (0.023) & 0.71 (2.08) 
\\
 \hline 
 & unpol  & 0.204 (0.144) & 0.195 (0.118) &3.67 (14.2) 
\\ 
 $ 4 j +2 \nu$  & pol $-$  & 0.294 (0.207) & 0.27  (0.156) &4.29 (16.5)  
 \\ 
                & pol $+$ & 0.113(0.080) &  0.126(0.077) & 3.08 (11.8)
\\
\end{tabular}
\end{center}
\caption{Values for the standard model, pure anomalous and
  interference cross sections (see Eq.\ \protect\ref{base})
for the $ZZZ$ production and several center--of--mass energies 
and $e^-$ polarizations.}
\label{sig:zzz}
\end{table}


\begin{table}
\begin{displaymath}
\protect
\begin{array}{||c|c|c|c||}
\hline\hline
\sqrt{s} \mbox{ GeV}& e^- \mbox{polarization (\%)}&\alpha_{4,5}  
& \alpha_{6,7,10}
\\
\hline \hline
 500   &           0    &   (-0.19,0.25)   & (-0.096, 0.12)
\\
 500   &          80    &   (-0.20,0.26)   & (-0.098,0.13)  
\\
 500   &          -80    &   (-0.18,0.22)   & (-0.088,0.11)  
\\
1000   &           0    &   (-0.14,0.15)   & (-0.068,0.075)
\\
1000   &          80    &   (-0.14,0.16)   & (-0.071,0.079)      
\\
1000   &          -80    &   (-0.12,0.13)   & (-0.058,0.063)      
\\
\hline\hline
\end{array}
\end{displaymath}
\caption{3$\sigma$ allowed values of the quartic anomalous couplings 
obtained from the reaction $e^+e^- \rightarrow ZZZ$. 
 }
\label{l:zzz}
\end{table}



\end{document}